\documentclass[twocolumn]{aastex62}
\usepackage[T1]{fontenc}
\usepackage[latin1]{inputenc}
\usepackage{enumerate,times,subfigure,multirow,xcolor} 
\usepackage{hyperref,astrojournals,amsmath,amssymb,mathtools,graphicx,txfonts,microtype,nicefrac,xspace}

\makeatletter

\newcommand{\Md}{M_{\rm disk}}
\newcommand{\Mbh}{M_{\rm \bullet}}
\newcommand{\Msun}{M_{\odot}}
\newcommand{\e}[1]{\times 10^{#1}}

\renewcommand*{\vec}[1]{\boldsymbol{#1}}
\newcommand{\dg}{^\circ}

\shorttitle{GR in ENDs}
\shortauthors{Wernke \& Madigan}

\begin{document}
\bibliographystyle{apj}

\title{The Effect of General Relativistic Precession on Tidal Disruption Events from Eccentric Nuclear Disks}

\author{Heather N. Wernke}
\affiliation{JILA and Department of Astrophysical and Planetary Sciences, CU Boulder, Boulder, CO 80309, USA}
\email{heather.wernke@colorado.edu}

\author{Ann-Marie Madigan}
\affiliation{JILA and Department of Astrophysical and Planetary Sciences, CU Boulder, Boulder, CO 80309, USA}

%%%%%%%%%%%%%%%%%%%%%%%%%%%%%%%%%%%%%%%%%%%%%%%%%%%%%%%%%%%%%%%%%%%%%%%%%%%%%%%%
\begin{abstract}
An eccentric nuclear disk consists of stars moving on apsidally-aligned orbits around a central black hole. The secular gravitational torques that dynamically stabilize these disks can also produce tidal disruption events (TDEs) at very high rates in Newtonian gravity. General relativity, however, is known to quench secular torques via rapid apsidal precession. 

Here we show that for a disk to black hole mass ratio of $\Md/\Mbh \gtrsim 10^{-3}$, the system is in the full loss cone regime. The magnitude of the torque per orbital period acting on a stellar orbit means that general relativistic precession does not have a major effect on the dynamics.  
Thus we find that TDE rates from eccentric nuclear disks are not affected by general relativistic precession.  Furthermore, we show that orbital elements between successive TDEs from eccentric nuclear disks are correlated, potentially resulting in unique observational signatures. \\ 
\end{abstract}


%%%%%%%%%%%%%%%%%%%%%%%%%%%%%%%%%%%%%%%%%%%%%%%%%%%%%%%%%%%%%%%%%%%%%%%%%%%%%%%%
\section{INTRODUCTION}  
\label{sec:intro}  

A tidal disruption event (TDE) occurs when a star is violently ripped apart by a black hole's tidal forces \citep{Hil75b}. When a star is tidally disrupted, roughly half of the stellar debris remains  
bound to the black hole while the other half of the debris escapes.  The gravitationally bound debris forms an accretion disk which feeds the black hole, producing a flare \citep{Ree88}. The current detection rate of flares from TDEs is about two per year \citep{vanVelzen2018} and this is expected to increase with new surveys such as the Large Synoptic Survey Telescope (LSST) \citep{vanVelzen2011}.
 
TDE flares can provide insight into the mysteries of many areas of astrophysics.  They illuminate central black holes in otherwise quiescent galaxies \citep{Maksym2013,Macleod2014}.  We can use their observations to test theories of accretion physics and relativistic jets \citep{Zauderer2011,Bloom2011,vanVelzen2016,Ale17}.  Tidal disruptions of white dwarfs should even produce gravitational waves detectable by the Light Interferometer Space Antenna (LISA) \citep{Macleod2014}.  Additionally, we can test our understanding of gravitational stellar dynamics near supermassive black holes by comparing theoretical TDE rates with observations.

\subsection{Loss cone dynamics} 
\label{subsec:loss-cone}

The rate of TDEs due to stellar two-body relaxation has been studied extensively \citep{Fra76,Lig77,Sha78,Coh78,Stone2016a}.  Two-body relaxation is the diffusive process by which stars exchange energy and angular momentum amongst themselves, sometimes scattering a star onto a tidally disrupting orbit.  It is faster to reach such an orbit by diffusion in angular momentum than in energy \citep{Fra76}. 

In order for a star in these systems to get close enough to the supermassive black hole (SMBH) to tidally disrupt, it must enter the loss cone.  
The loss cone defines the region containing orbits with pericenters inside the tidal disruption radius of the black hole.  The tidal disruption radius is:
\begin{equation} \label{eq:rt}
r_t = \left(\frac{\Mbh}{M_*}\right)^{1/3}R_*,
\end{equation}
where $\Mbh$ is the mass of the black hole, $M_*$ is the mass of the star, and $R_*$ is the radius of the star \citep{Ree88}.  Orbits within the loss cone have angular momenta less than the angular momentum of an orbit with a pericenter equal to the tidal radius,
\begin{equation} \label{eq:losscone}
J < J_{LC}\approx \sqrt{2G\Mbh r_t}.
\end{equation} 
There are two loss cone regimes, defined by the parameter q, 
\begin{equation} \label{eq:qparam}
q=\left(\frac{\Delta J_P}{J_{LC}}\right)^2,
\end{equation}
where $\Delta J_P$ is the change in angular momentum per orbital period \citep{Lig77}.  If $q\ll 1$, stars take multiple orbital periods to enter the loss cone.  This is known as the empty loss cone regime or the diffusion limit because the time for a star to enter the loss cone is greater than the time for the star to be destroyed.  If $q\gg 1$, stars can jump into and out of the loss cone within one orbital period.  This is known as the full loss cone regime or the pinhole limit, because the loss cone is continuously populated by stars.
The division between the two loss cone regimes for a spherical nuclear star cluster lies close to the radius of influence of the black hole \citep{Lig77}. The TDE rate, in this case, is also dominated by stars coming from this region. 

\subsection{Status of observations of TDEs} 
\label{subsec: obs-tdes}

In deriving theoretical TDE rates, we typically assume that stars come from an isotropic, spherical distribution around the black hole and are driven to the black hole through two-body relaxation \citep{Wan04,Stone2016a}.  Theoretical TDE rates in these spherical nuclear star clusters have been calculated to be $2.1\e{-4}$ yr$^{-1}$ gal$^{-1}$ \citep{Wan04}, and more recently $2.9\e{-5}$ yr$^{-1}$ gal$^{-1}$ \citep{Stone2016a}.  In observations, however, TDEs are preferentially found in post-merger or post-starburst galaxies (K+A/E+A galaxies) at much higher rates. K+A/E+A galaxies are a relatively rare subtype of elliptical galaxy that underwent a major starburst about 1-1.5 Gyr ago \citep{Couch87,Pog04}.  K+A/E+A galaxies make up $0.2\%$ of the galaxies in the local universe, and yet, the observed TDE rates in these K+A/E+A galaxies are $1-3\e{-3}$ yr$^{-1}$ gal$^{-1}$, which pushes the observed TDE rate of `normal' galaxies down to $1-5\e{-6}$ yr$^{-1}$ gal$^{-1}$ \citep{French2016}.  There is even (tentative) evidence that the TDE rate could be as high as $10^{-1}$ yr$^{-1}$ gal$^{-1}$ in ultra-luminous infrared galaxies (ULIRGs), which are typically in the process of merging \citep{Tadhunter2017,Dou2017}.  We learn from these observations that merging galaxies and post-merger galaxies tend to have elevated TDE rates.

Several dozen TDE candidates have been identified in the last two decades, from UV/optical to X-ray.  TDE candidates are generally identified as flaring events, inconsistent with supernovae, at the centers of galaxies.  Candidates are typically excluded if the host galaxy shows signs of AGN activity. 
There have been a number of alternative ideas to explain these flaring events at galactic centers. Proposed TDE impostors include supernovae in AGN disks and black hole accretion disk instabilities \citep{Sax16}.  One distinguishing feature that can be used to discriminate between real TDEs and impostors is the critical black hole mass beyond which a TDE will not be observable, known as the Hills mass \citep{Hil75b}.  The Hills mass results from the fact that the tidal radius and Schwarzschild radius of a black hole scale differently with mass of the black hole. The Schwarzschild radius is given by
\begin{equation}
r_s = \frac{2 G \Mbh}{c^2}, \label{eq:rsM}
\end{equation}
where $G$ is the gravitational constant and $c$ is the speed of light.   
Equating the tidal radius to the Schwarzschild radius yields a Hills mass of $\sim$$10^8\Msun$ for a solar-type star.
Above this limit, the star plunges into the black hole without emitting a flare.  A rapidly spinning black hole can raise this limit to $\sim$$10^9 \Msun$ \citep{Kesden2012}. Recently, \citet{vanVelzen2018} presented the black hole mass function of optical/UV-selected TDE candidates and showed a sharp decrease in the number of candidates above $\Mbh = 10^{7.5}\Msun$.  This is consistent with the direct capture of stars when the black hole is above the Hills mass and provides strong evidence that we are seeing TDEs rather than impostors.  

\subsection{Secular dynamics and eccentric nuclear disks} 
\label{subsec:ends-and-sec}

Two-body relaxation is not the only form of relaxation present in galactic nuclei. Resonant relaxation\footnote{Note that this is a secular (orbit-averaged) effect; `resonant' here refers to the resonance between the azimuthal and radial frequency of a Kepler orbit.} arises in near-Keplerian potentials \citep{Rauch1996}.  
A particle on a near-Keplerian orbit traces out the same path repeatedly.   
On a timescale less than the precession timescale, the orbits remain $\sim$fixed, and exert mutual gravitational torques on each other. Thus, the angular momentum relaxation can be greatly enhanced, while the energy relaxation is unaffected \citep{Rauch1996}. 

Resonant relaxation is most effective for stars orbiting close to the central supermassive black hole (in the absence of general relativity).  
This means that in an isotropic, spherical stellar distribution, where TDEs come most often from near the radius of influence, resonant relaxation will not greatly increase the rate or number of TDEs \citep{Rau98}.  Not all galactic nuclei, however, are $\sim$spherical like our galactic center.  The nucleus of our nearest galactic neighbor, Andromeda (M31), has a very different configuration. 

The Andromeda Galaxy (M31) has an elongated nucleus that resolves into two distinct brightness peaks.  
The double-nucleus can be explained by a thick, apsidally-aligned eccentric nuclear disk of Keplerian orbits around a SMBH  \citep{Tremaine1995}.  
The two brightness peaks correspond to apoapsis and periapsis of the eccentric nuclear disk. 

While it may seem like the central disk in M31 is an unusual and unlikely arrangement, the fact that we see it in our closest major galaxy suggests that it is a common configuration.  In fact, despite observational challenges, \citet{Lauer2005} found that about 20\% of nearby, early-type galaxies have features consistent with eccentric nuclear disks seen from different angles on the sky.

\subsection{TDEs from eccentric nuclear disks} 
\label{subsec:tdes-ends}
The stability of eccentric nuclear disks has long been a mystery.  One would expect that the apsidal precession of individual orbits would spread out the disk into an axisymmetric structure on a timescale much shorter than the age of the stars.  In a recent paper \citep{Madigan2018}, we proposed that the same secular mechanism that stabilizes eccentric nuclear disks is responsible for producing high rates of TDEs. 
  
The forces that cause precession in eccentricity vectors also result in a build-up of gravitational torques between orbits. 
These torques change the eccentricities of individual orbits as they are perturbed ahead of, or behind, the disk. Differential precession driven by these eccentricity changes holds the disk together.  
The orbits in an eccentric nuclear disk
undergo oscillations in eccentricity.  
During the high eccentricity phase of an oscillation, a star 
can be tidally disrupted as it moves through pericenter.  The gravitational torques due to secular dynamics are much more efficient at refilling the loss cone than two-body relaxation, which has typically been used to determine TDE rates.  We proposed that secular torques in eccentric nuclear disks can produce the observed high rate of TDEs in K+A/E+A galaxies \citep{Madigan2018}.  \citet{Hop10a,Hop10b} show that eccentric nuclear disks can form via the merging of gas-rich galaxies, meaning that it would be likely to find eccentric nuclear disks in post-merger, K+A/E+A galaxies.  TDE rates in eccentric nuclear disks could be as high as $\sim 1$ yr$^{-1}$ gal$^{-1}$ at early times in the life of the disk \citep{Madigan2018}.

Several other mechanisms have been theorized to explain the enhanced TDE rates in K+A/E+A galaxies.  One of these theories is an enhanced rate due to SMBH binaries after the starburst.  \citet{Chen2011} show that the TDE rate should scale weakly with the SMBH mass ratio.  This would indicate that TDEs would be seen primarily after minor mergers, which are more common.  TDEs are preferentially observed, however, in mergers with a more equal SMBH mass ratio, indicating that the TDE rate is not driven by SMBH binaries \citep{French2017}.  Another theory involves more dense spherical star clusters resulting in enhanced two-body relaxation \citep{Stone2016b,Stone2017}.  

\subsection{This work} 
\label{subsec:this-paper}

In \citet{Madigan2018}, we evolved eccentric nuclear disks with $N$-body simulations in Newtonian gravity. Rapid apsidal precession due to general relativity, however, can quench secular dynamical mechanisms; a well-known example of this is the Kozai-Lidov effect \citep{Ford2000,Blaes2002,Naoz2013b}.  Resonant relaxation in a spherical cluster also gets quenched at low semi-major axes by general relativistic precession as the orbits move too rapidly to allow torques to build up coherently \citep{Rauch1996,Madigan2011}.

Similarly, one might expect general relativistic precession to disrupt the secular torques of the eccentric nuclear disk, greatly decreasing the TDE rate.  As eccentricity increases due to secular torques, the general relativistic precession rate also increases (Equation \ref{eq:precfreq}).  One would therefore expect eccentric orbits to precess ahead of the disk, escaping completely until joining back up on the other side and re-circularizing.  In this case, we should see fewer TDEs with general relativity than without it.

The goal of this work is to explore the effects of general relativity on TDEs occurring in eccentric nuclear disks, and to quantify the distribution of orbital elements of TDEs that originate in eccentric nuclear disks.  We do this using $N$-body simulations with and without general relativity.  We present the paper in the following manner: in Section \ref{sec:sims} we describe the initial conditions and parameters for our simulations, and compare the number of TDEs that occur with and without general relativity. We track the orbital elements of a single tidally disrupted star in order to show how quickly the orbit is torqued to an extreme eccentricity.  In Section \ref{sec:uoe} we explore the unique orbital elements of tidally disrupted stars from eccentric nuclear disks, including the penetration factor, inclination distribution, and change in eccentricity vector between TDEs.  In Section \ref{sec:disc} we summarize and discuss our results.

\section{$N$-body Simulations of Eccentric Nuclear Disks with General Relativistic Precession} 
\label{sec:sims}

We run $N$-body simulations of eccentric nuclear disks with {\tt REBOUND} \citep{Rein2012} and the {\tt IAS15} integrator \citep{Rein2015}.  We implement general relativity as a post-Newtonian approximation with {\tt REBOUNDX}\footnote{https://github.com/dtamayo/reboundx}.    
In this paper, we show results from simulations with the following parameters: N=100 stars\footnote{In \citet{Madigan2018}, we used a range of $N=100-1000$ stars.  Our results were qualitatively the same for the different $N$.  We use $N=100$ stars in order to reduce computing time.}, each with an initial eccentricity of 0.8, a range of semi-major axes ($a=1-2$) with a surface density of $\Sigma\propto a^{-2}$, Rayleigh distributed inclinations with mean $0.1\dg$, and a disk mass of $10^{-2}\Mbh$.  We want to qualitatively understand the effects of general relativistic precession rather than obtain an exact number for the TDE rate.

In each of these simulations, we examine the effect that general relativity has on the number of tidal disruption events.    
The orbit-averaged precession rate due to general relativity is given by  
\begin{equation} \label{eq:precfreq}
\dot{\omega}_{GR}=\frac{6\pi G\Mbh}{ac^2\left(1-e^2\right)}.
\end{equation}
Equation \ref{eq:precfreq} is a first order post-Newtonian approximation in general relativity yielding corrections to Newtonian accelerations of $\mathcal{O}(v^2/c^2)$ \citep{Ein16}.   
We track the general relativistic precession rate in our simulations by calculating the change in the orientation of the eccentricity vector at each time step.  

A star is considered tidally disrupted if at any point in the simulation  its radius $\leq r_t$.  
We treat stars as point masses and do not extract them from our simulation after they are disrupted, but they are counted only once as a TDE.

\subsection{Effects of General Relativity} 
\label{sec:gr}

\begin{figure}[ht!]
\centering
\includegraphics[trim=7cm 9cm 7cm 9cm, clip=true, width=\columnwidth]{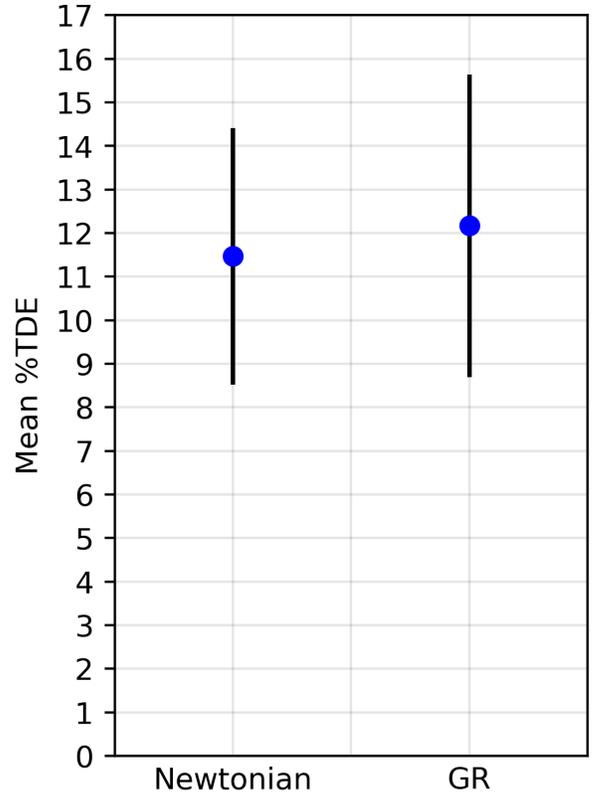}
\caption{\textbf{Mean percent of disk stars tidally disrupted in Newtonian gravity and with general relativistic precession.}  Here we see that eccentric nuclear disks with Newtonian gravity and with general relativistic precession each have a mean of roughly 12\% of disk stars that tidally disrupt.  The error bars show the standard deviation.  By performing a student's t-test for statistical significance, we find that these are not significantly different.} 
\label{fig:percent}
\end{figure}

We find that the TDE rate with general relativistic precession is the same as in Newtonian gravity.  About 12\% of disk stars are tidally disrupted\footnote{This percentage is smaller than in \citet{Madigan2018} because we have a more rigorous TDE criterion and exclude partial disruptions from our analysis.}  for a $10^6\Msun$ black hole during a time of 1000 orbital periods, where each orbital period is roughly 1000 years.  
We have compiled results from $\sim$45 simulations with general relativistic precession and $\sim$100 simulations without general relativistic precession.  The mean percent and standard deviation of tidally disrupted disk stars is shown in Figure \ref{fig:percent}.  The number of TDEs is approximately equal for both general relativistic simulations and Newtonian simulations.  This means that general relativistic precession does not quench secular torques in eccentric nuclear disks.

\begin{figure*}[tb]
\centering
\includegraphics[trim=2cm 4cm 2cm 4cm, clip=true, width=\textwidth]{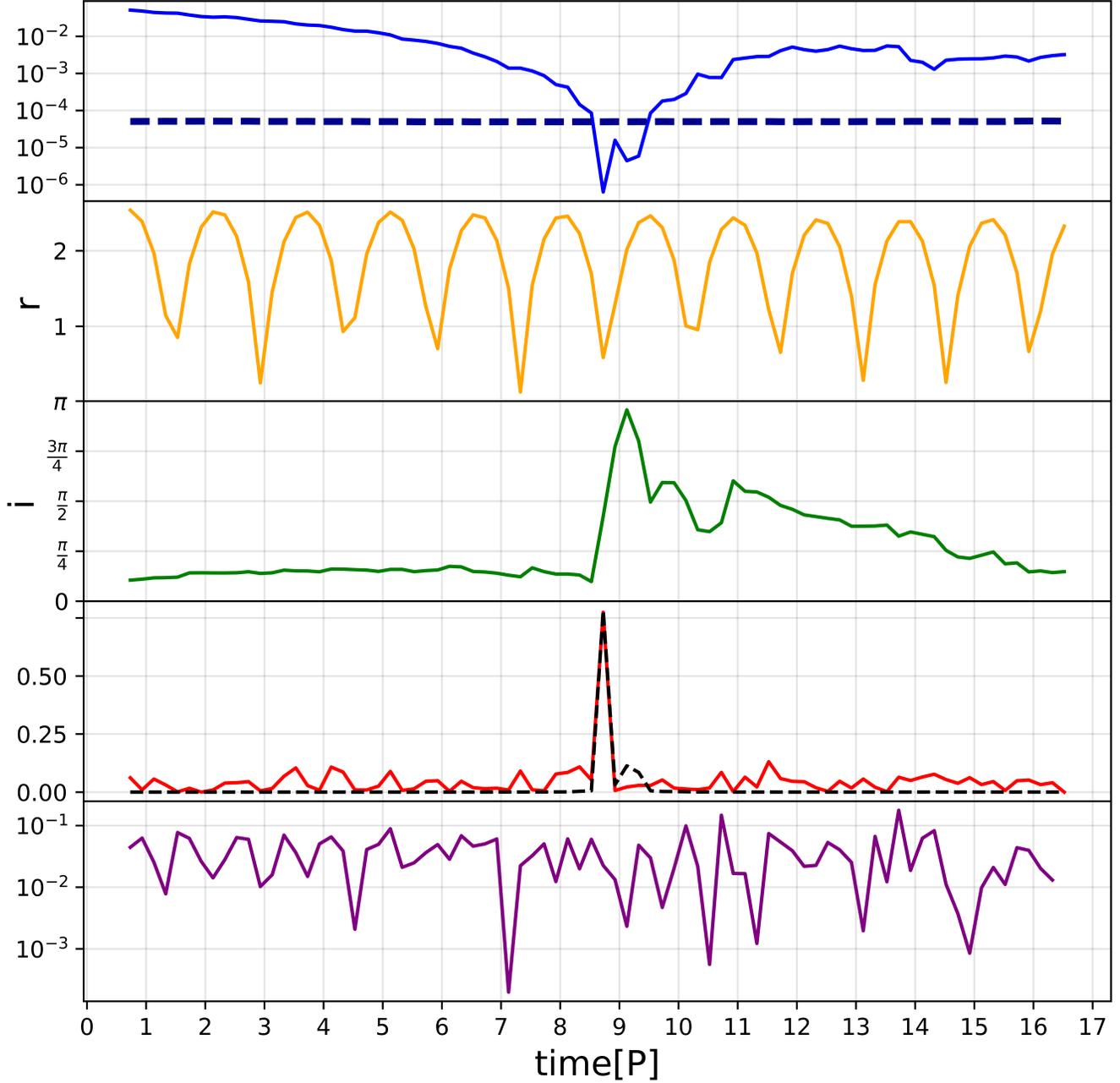}
\caption{\textbf{A tidal disruption event (TDE) in an N-body simulation of an eccentric nuclear disk with general relativistic precession.}  The top panel shows the eccentricity of a star undergoing a tidal disruption event as a function of time in units of orbital periods ($P$).  At the time of tidal disruption, the eccentricity increases from $\sim$0.999 to $>0.9999$ within one orbital period.  The dashed line shows the $1-e$ value such that $J=J_{LC}$.  The second panel shows the orbital radius (in code units) of the same star as a function of time.   
The star meets the requirement of being at pericenter to be tidally disrupted.   
The third panel shows that the inclination of this star flips by $180\dg$ at the same time it reaches an extreme eccentricity.  The fourth panel shows the general relativistic precession rate with the time derivative of the eccentricity vector $(i_e')$ in units of radians per orbital period.  The black dashed line shows the analytic general relativistic precession rate $(\dot{\omega}_{GR})$ from Equation \ref{eq:precfreq}.  General relativistic precession is effective for less than one orbital period, allowing the star to still tidally disrupt.  The normalized torque required to produce a TDE is on the order of $10^{-2}$ per orbital period.  In the last panel, we see that the normalized torque applied to the orbit of the star oscillates around $10^{-2}$ per orbital period, ensuring that even with general relativity, the tidal disruption event can occur.}
\label{fig:precess}
\end{figure*}

In order to understand this, we track the orbital elements of a single star (with general relativistic precession) which suffers a tidal disruption event in Figure \ref{fig:precess}.  
We see a star that develops an eccentricity such that its orbital angular momentum is less than the loss cone angular momentum.  The star also passes through pericenter while it is at a high eccentricity, meaning that the star is close enough to the black hole to be tidally disrupted.  We also see that the star's orbital inclination flips by $\sim$$180^\circ$ as it reaches extreme eccentricity (see discussion in Section \ref{sec:incdist}).  Panel 4 shows the general relativistic precession rate, which we track by calculating the change of $i_e$ in each time step.  $i_e$ tracks the orientation of the eccentricity vector in the plane of the disk and is given by
\begin{equation} 
i_e=\arctan\left(\frac{e_y}{e_x}\right),
\label{eq:ie}
\end{equation}
\citep{Madigan2016}.  Here $e_x$ and $e_y$ are the x and y components of the eccentricity vector. 
We use $i_e$ instead of the argument of periapsis, $\omega$, or the longitude of periapsis, $\varpi$, to avoid effects of changing inclination.  As an orbit rolls over its major axis, the eccentricity vector remains close to the $x-y$ plane, even though the inclination grows.  $\omega$ and $\varpi$, however, will change with the flipping inclination.  We see that the rate of change of $i_e$ is very small until the star reaches pericenter at an extreme eccentricity where there is a large jump due to general relativity.  This jump in precession rate is only present for a fraction of an orbital period.  The final panel of Figure \ref{fig:precess} shows the torque acting on the orbit in units of the circular angular momentum which we explore in the next section. 

\subsection{Magnitude of Torque from Disk} \label{sec:torque}

Here we calculate the magnitude of the torque exerted on a typical orbit by the disk.  The orbit is described by its specific angular momentum and energy
\begin{subequations}
\begin{align}
j^2 &=  G\Mbh a ~(1 - e^2) \label{eq:j} \\
E &= \frac{G\Mbh}{2 a}. \label{eq:E}
\end{align}
\end{subequations}
For an eccentric orbit, the specific torque is given by
\begin{equation} \label{eq:torque}
\vec{\tau}=\vec{j}^\prime=\vec{r}\times \vec{f}.
\end{equation}
$\vec{r}$ is the orbital radius and $\vec{f}$ is the specific gravitational force felt by an orbit due to the rest of the disk.  This force is defined by
\begin{equation}
\left|\vec{f}\right|=\frac{G\Md}{r^2},
\end{equation}
where $\Md$ is the mass of the eccentric nuclear disk.  Approximating $r$ by the semi-major axis $a$ yields a torque
\begin{equation}
\left|\vec{\tau}\right|\approx\frac{G\Md}{a}.
\end{equation}
Normalizing the torque by the circular angular momentum $\left(j_c = \sqrt{G\Mbh a}\right)$ yields
\begin{equation} \label{eq:torqueoverJc}
\left|\frac{\vec{\tau}}{j_c}\right| \approx \frac{\Md}{\Mbh} \frac{2 \pi}{P},
\end{equation}
where $P = 2\pi\sqrt{{a^3}/{G\Mbh}}$ is the orbital period. 
Hence, in our $N$-body simulations, in which ${\Md}/{\Mbh} = 10^{-2}$, the normalized torque per orbital period should be on the order of $6 \times 10^{-2}$. 
The final panel in Figure \ref{fig:precess} shows that indeed our example star experiences a torque of $\mathcal{O}(\rm{few} \times 10^{-2})$.
This magnitude of torque can change an orbit's eccentricity from $e\approx{0.998}$ to $e\approx{1}$ within one orbital period. That is, the change in angular momentum required to produce a TDE can occur within one orbital period, suggesting that our system is in the full loss cone regime.  By assuming a $10^6\Msun$ black hole and a solar-type star, we find from Equation \ref{eq:qparam}, that $q\approx 40$ in our simulations, putting the system well within the full loss cone or pinhole regime.  
This explains why general relativity is ineffective at shutting down the TDE production. In the full loss cone regime, a stellar orbit can be propelled from outside the loss cone to inside in less than an orbital period. General relativistic precession only acts strongly when the star approaches pericenter, at which point it is too late to avoid disruption.

Not all eccentric nuclear disks will be in the full loss cone regime. 
The transition from full loss cone to empty loss cone occurs when $q=1$ such that 
\begin{equation}
\frac{\Md}{\Mbh} = \sqrt{\frac{r_t}{2 \pi^2 a}}.
\end{equation}
For a SMBH of $10^6\Msun$, solar-type stars, and a disk inner edge of $a=0.05$ pc, we find that disks with 
$\Md/\Mbh \geq 1.5\e{-3}$ are in the full loss cone regime.

\section{UNIQUE ORBITAL ELEMENTS}
\label{sec:uoe}

Two-body relaxation predicts that the time between individual TDEs ($\sim$$10^4$ years) is much greater than the time it takes for a TDE disk to accrete onto the black hole. 
If stars come from eccentric nuclear disks however, the typical timescale between individual TDEs can be much shorter \cite[$\sim$1-10 yr;][]{Madigan2018}, and TDE disks could potentially overlap with one another.  This could have interesting observational consequences especially if the orbital parameters of TDEs are correlated.

\subsection{Penetration Factor}
\begin{figure}[ht!]
\centering
\includegraphics[trim=3.5cm 9cm 3.5cm 9cm, clip=true, width=\columnwidth]{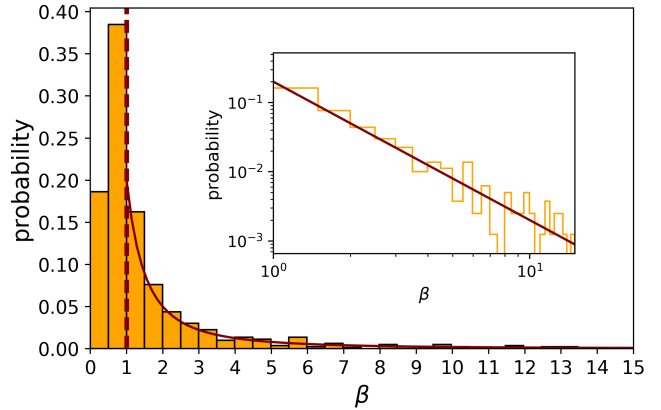}
\caption{\textbf{The distribution of impact parameters with general relativity.}  The dashed line shows the critical value of the impact parameter.  A star with an impact parameter less than one will be tidally squeezed and stretched (many will even lose mass), but not fully disrupted.  A disruption event with an impact parameter greater than 23.5 (corresponding to solar-type stars and a non-spinning $10^6\Msun$ black hole) will not be visible to observers because its pericenter is inside the Schwarzschild radius.  We see in this figure that our results are consistent with the pinhole regime in which orbits have large steps in angular momentum, allowing stars to jump into the loss cone within an orbital period.  The probability distribution function of the impact parameter is well-fit by the curve $\propto\beta ^{-2}$, as shown by the solid maroon line.  The inset shows the same histogram with the curve $\propto\beta ^{-2}$ on a log-log scale.}
\label{fig:betadist}
\end{figure}

The strength of a tidal disruption may be quantified by the dimensionless penetration factor,
\begin{equation}
\label{eq:beta}
\beta = \frac{r_t}{r_p},
\end{equation}
where $r_t$ is the tidal radius and $r_p$ is the pericenter of the star's orbit \citep{Pre77}.  In Figure \ref{fig:betadist} we show the distribution of penetration factors in our simulations.  
If the penetration factor is greater than or equal to one, the star will be tidally disrupted. If the penetration factor is less than, but close to one, the star may have its outer layers stripped, with a stellar core remaining intact \citep{Iva01,Guillochon2013,Bog14,Mai17}.  If the penetration factor is too large however, the star will fall straight into the black hole without emitting an electromagnetic flare. This occurs when the $r_p < r_s$. 
For a non-spinning, $10^6\Msun$ black hole and solar-type stars this occurs at $\beta = 23.5$.  

We see in Figure \ref{fig:betadist} that the probability distribution function, $P\propto \beta^{-2}$, is fully consistent with the full loss cone or pinhole regime \citep{Coughlin2017}. This is significant because the critical radius (where $q=1$) is typically found near the radius of influence of the black hole.  We find that eccentric nuclear disks bring the critical radius orders of magnitude within the radius of influence, to a radius smaller than the inner edge of the disk.

\subsection{Inclination Distribution}
\label{sec:incdist}

In a spherical, isotropic stellar system dominated by two-body relaxation, there should be no correlation between the orbital angular momentum vectors of consecutive TDEs, and so we would expect to see an isotropic distribution of TDE inclinations. This is quite different in the case that stars are originating in an eccentric nuclear disk.

\begin{figure}[th!]
\centering
\includegraphics[trim=3cm 9cm 3cm 9cm, clip=true, width=\columnwidth]{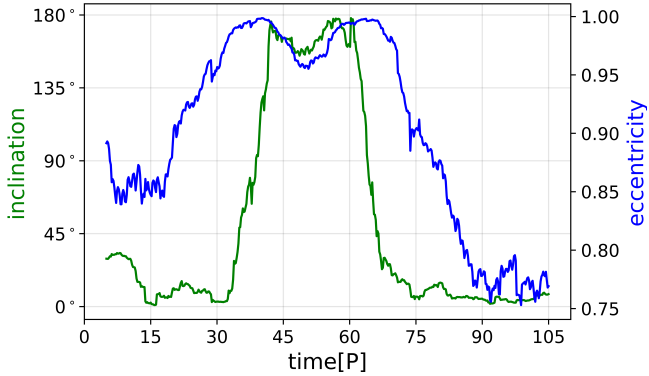}
\caption{\textbf{The inclination flip of a star at high eccentricity.}  The blue line shows the eccentricity of the star, while the green line shows the inclination of the orbit in radians.  The inclination flips from $0\dg$ to $180\dg$ or vice versa, corresponding to the extreme peaks in eccentricity.  An orbit with an inclination between $0\dg$ and $90\dg$ is prograde and an orbit with inclination between $90\dg$ and $180\dg$ is retrograde.}
\label{fig:incflip}
\end{figure}

We find that whenever a star reaches a high eccentricity in our simulations, it undergoes an inclination flip of $180^\circ$.  Figure \ref{fig:incflip} is an example of a double peak in eccentricity corresponding to a double $180^\circ$ flip in inclination.  As the orbit is negatively torqued by the disk to extreme eccentricity (in blue), its angular momentum vector decreases until it passes through zero. 
At this point, the inclination (in green) flips $180^\circ$ and the orbit switches from a prograde orientation (with respect to the disk) to a retrograde orientation.  The orbit now feels a positive torque causing it to circularize and precess quickly back towards the disk. On the other side of the disk, the angular momentum will again decrease, pass through zero and change direction.  The orbit is prograde again after the second flip.  
These double peaks of inclination were also seen in our Newtonian simulations \citep{Madigan2018}. 

\begin{figure}[th!]
\centering
\includegraphics[trim=2cm 10.5cm 2cm 10.5cm, clip=true, width=\columnwidth]{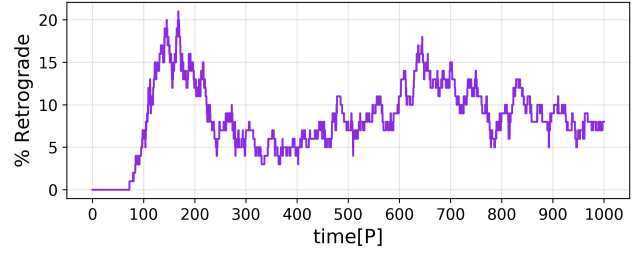}
\caption{\textbf{The percentage of disk stars on retrograde orbits with time.}  A typical plot of the changing percentage of retrograde orbits with time for a single simulation.  The eccentric nuclear disk simulation begins with all orbits in a prograde orientation.  The number of retrograde orbits climbs quickly after about 150 orbital periods to a peak of $\sim$20$\%$ retrograde and then oscillates around 10\% retrograde for the rest of the simulation.}
\label{fig:percentretro}
\end{figure}

As the orbits flip from prograde to retrograde and back, the percentage of stars on retrograde orbits fluctuates throughout a given simulation.  Figure \ref{fig:percentretro} is an example of the percentage of retrograde orbits in a disk for a single simulation. Most of these retrograde orbits lie near the inner edge of the disk. This may lead to interesting observational signatures in the velocity moments of eccentric nuclear disks. 

\begin{figure}[ht!]
\centering
\includegraphics[trim=3.25cm 9cm 3.25cm 10cm, clip=true, width=\columnwidth]{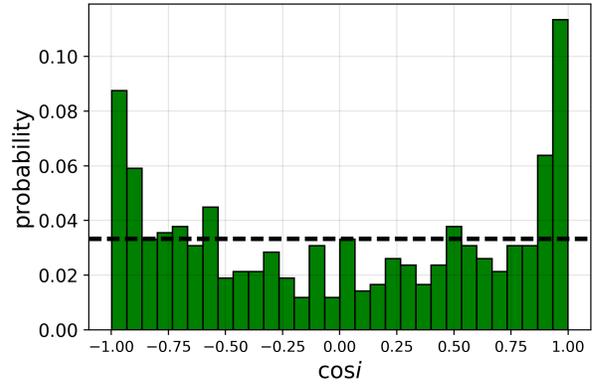}
\caption{\textbf{The distribution of orbital inclinations of disrupted stars with general relativity.}   
The black dashed line shows the isotropic distribution of cosine of inclinations that we would see from a spherical cluster. Stars originating in an eccentric nuclear disk preferentially tidally disrupt at inclinations of $0\dg$ and $180\dg$.}
\label{fig:cosincdist}
\end{figure}

The flipping of orbits in inclination results in an anisotropic distribution of TDE inclinations (see Figure \ref{fig:cosincdist}).
Stars preferentially tidally disrupt at orbital inclinations of $0\dg$ and $180\dg$ with respect to the disk mid-plane. More disruptions occur at $0\dg$. 
This is because the stars get the first opportunity to disrupt at an inclination of $0\dg$, while their orbit is ahead of the disk. 
The probability for a star to disrupt in one orbital period is 
\begin{equation}
\begin{aligned}
P_{\rm TDE} &=\frac{J_{LC}}{\Delta J_P} \\
        &= \frac{1}{\sqrt{2}\pi} \sqrt{\frac{r_t}{a}} \left(\frac{\Mbh}{\Md}\right).
\end{aligned}
\end{equation}
We estimate $a \approx 10^{-2}~r_H$ for the inner edge of the disk, where $r_H$ is the radius of influence of the black hole, based on the disks in M31 and the Galactic center \citep{Madigan2018}.   
We take $r_H$ to be 5pc from observations of the Galactic center \citep{Lu2009}. 
We find that $P_{\rm TDE}=0.15$.  Out of 100 stars vulnerable to disruption, $\sim$15 will tidally disrupt at an inclination of $0\dg$.  85 stars will then flip inclinations and have a 15\% chance ($\sim$12-13) of tidally disrupting at an inclination of $180\dg$. Therefore, we find that the number of TDEs at $180\dg$ is 85\% the number at $0\dg$, or in general,
\begin{equation}
N_{\rm TDE ~(i=180\dg)}= (1 - P_{\rm TDE}) ~ N_{\rm TDE ~(i=0\dg)}.
\end{equation}
This explains the height difference that we see in the inclination distribution in Figure \ref{fig:cosincdist}.

For a spinning Kerr black hole, the tidal and capture cross-sections shift towards negative angular momenta \citep{Bel92}. The asymmetric cross-sections make it easier for stars on retrograde orbits to be captured, meaning that prograde TDEs will be preferentially observed.        
If the spin angular momentum vector of the black hole is aligned with the orbital angular momentum vector of the disk, then the preference for TDEs from eccentric nuclear disks to have $\sim$$0\dg$ orbital inclination puts them at the perfect orientation to be visibly disrupted by a Kerr black hole. 

\subsection{Eccentricity Vector}

\begin{figure}[ht!]
\centering
\includegraphics[trim=3.5cm 7cm 3.25cm 7cm, clip=true, width=\columnwidth]{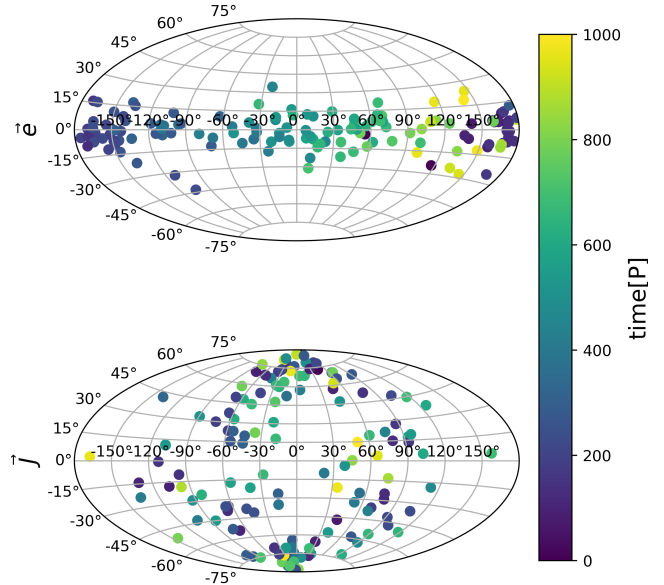}
\caption{\textbf{Aitoff projection of eccentricity and angular momentum vectors of TDEs.} Colors indicate the time of TDE in units of orbital periods at the inner edge of the disk. The eccentricity vectors, $\vec{e}$, precess together in a prograde direction, staying close to the mid-plane.  The orbits start with their eccentricity vectors at about $60\dg$ and then precess together in a prograde direction (to the right on our projection plot).  The first TDEs begin occurring after about 200 orbital periods, when the disk has precessed to $\sim$$0\dg$.  When the orbits flip in inclination, they flip over the major axis instead of the latus rectum.  The spread in the angular momentum vectors, $\vec{j}$, confirms that the orbits roll over their major axes.}
\label{fig:e_aitoff}
\end{figure}

\begin{figure}[ht!]
\centering
\includegraphics[trim=6cm 9.5cm 6cm 9.5cm, clip=true, width=\columnwidth]{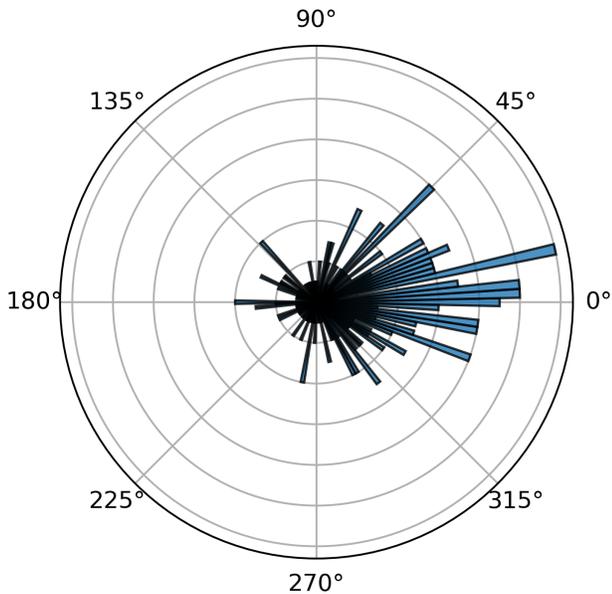}
\caption{\textbf{The distribution of $\Delta \theta$ between TDEs.}  A histogram of $\Delta \theta$ for consecutive TDEs with general relativistic precession.  About 20\% of the consecutive TDEs occur with a small ($\leq 20\dg$), positive change in $i_e$. This is partially due to the prograde precession of the disk.  The clustering of $\Delta \theta$ around $0$ and slightly greater than $0$ show that the first condition for tidal streams crossing is met for many consecutive TDEs.  Prograde precession is in the positive (counter-clockwise) direction.}
\label{fig:die}
\end{figure}

We plot the eccentricity vectors of tidally disrupted stars at the time of TDE in the top panel of Figure \ref{fig:e_aitoff}.  The eccentricity vectors precess together in a prograde direction while remaining in the plane.  This means that when a stellar orbit flips in inclination, it flips over its major axis.  The bottom panel of Figure \ref{fig:e_aitoff} shows the angular momentum vectors of the same tidally disrupted stars at the time of TDE.  The spread in angular momentum vectors confirms that the orbits roll over their major axes.

With TDEs preferentially occurring in the plane with inclinations of $0\dg$ or $180\dg$, the debris streams from two sequential TDEs could cross and produce unique observational signatures.  \citet{Bonnerot2018preprint} derive the conditions necessary for a tidal stream crossing to occur, which depend on disk properties, tidal stream widths, and the time between consecutive TDEs.   
If the pericenter shift between tidally disrupted stars is positive, the time delay is small, and the inclination offset is less than the width of the tidal streams, crossing of the tidal streams could occur.  While our simulations do not allow for the calculation of accurate time delays (due to the low $N$ nature of our simulations) or stream width, we calculate the pericenter shift between two TDEs with an angle, $\Delta \theta$.   
This $\Delta \theta$ tracks the orientation of the orbit of the first TDE with respect to the orientation of the orbit of the second TDE.  We track the orientation of the orbits in our simulation with $i_e$, defined in Equation \ref{eq:ie}.  We, therefore, calculate $\Delta \theta$ between TDEs as
\begin{equation}
\label{eq:dtheta}
\Delta \theta = i_{e_2}-i_{e_1}.
\end{equation}

We show, in Figure \ref{fig:die}, a distribution of the $\Delta \theta$ between pairs of TDEs in our simulations.  We see that about 20\% of consecutive TDEs occur with a small ($\leq 20\dg$), positive $\Delta \theta$, satisfying one of the conditions for tidal streams crossing. We expect that this condition would be met due to prograde precession of the eccentric nuclear disk. 

\section{DISCUSSION}
\label{sec:disc}

This paper focuses on the dynamics of eccentric nuclear disks with general relativistic precession. 
In \citet{Madigan2018}, we showed that the same secular mechanism that keeps eccentric nuclear disks stable results in extremely high TDE rates.  This work did not include general relativity, however, which is known to quench secular torques via rapid apsidal precession.  In this paper, we show that secular gravitational torques push the orbits of stars to extremely high eccentricities within one orbital period (full loss cone regime).  This does not allow general relativistic precession enough time to suppress the TDE rate.  
The geometry of eccentric nuclear disks is key: the torques acting on an orbit from the rest of the disk stars are coherent. Our results point to the following conclusions and implications:
\begin{enumerate}

\item
General relativistic precession does not affect the TDE rate from eccentric nuclear disks as stars at the inner edge of the disk are in the full loss cone regime.  TDEs occur in simulations with general relativity as often as they occur in Newtonian simulations. 

\item
TDEs from eccentric nuclear disks do not follow an isotropic distribution of inclinations; they preferentially disrupt at inclinations of $0\dg$ and $180\dg$ with respect to the mid-plane of the disk. Overlapping TDE disks may have similar (or opposing) angular momenta that can build up (or cancel each other out).

\item
The probability of disrupting stars on prograde orbits is higher than the probability of disrupting stars on retrograde orbits for a non-spinning, Schwarzschild black hole in an eccentric nuclear disk.  Similarly, the number of prograde captured stars (within the Schwarzschild radius so that a flare will not be observed), will also be greater than the number of retrograde captured stars for a non-spinning, Schwarzschild black hole. 

\item
If an eccentric nuclear disk forms during a gas-rich major merger, it is likely that the central gaseous accretion event that produces the disk aligns the disk angular momentum with that of the central SMBH. 
Spinning, Kerr black holes have asymmetric tidal and capture cross-sections \citep{Bel92}.  For black holes with mass greater than the Hills mass, the only observable TDEs are those on prograde orbits aligned with the black hole spin. The preference for TDEs from eccentric nuclear disks to have $\sim$$0\dg$ orbital inclination puts them in the perfect orientation to be observably disrupted by such black holes.  There may be evidence of TDEs by extremely massive black holes already.  
ASSASN-15lh is a TDE candidate found in a galaxy with a central SMBH much more massive than the Schwarzschild Hills mass ($\sim$$10^{8.24} \Msun$) \citep{Leloudas2016}.  

\item In steady-state, eccentric nuclear disks have a non-negligible fraction of retrograde orbiting stars ($\sim$10\%). Most of these will lie at the inner edge of the disk. This will lead to interesting observational signatures in the velocity moments of eccentric nuclear disks. 

\end{enumerate}

Finally, we look back to our nearest neighbor, Andromeda (M31).  To date, no TDEs have been observed from its center.  This may be due to the fact that the eccentric nuclear disk in M31 is very old, on the order of Gyr \citep{Sil98}.
Unless continuously replenished, an eccentric nuclear disk loses mass due to stars being destroyed by tidal forces, but it does not lose significant angular momentum.  The disk, therefore, becomes less eccentric with time, causing the TDE rate to decrease \citep{Madigan2018}. At $\Mbh \simeq1.4\e{8}\Msun$, the mass of the M31 black hole is also greater than the Hills mass \citep{Bender2005}. We should not expect to observe TDEs, unless the black hole is spinning. \\

%%%%%%%%%%%%%%%%%%%%%%%%%%%%%%%%%%%%%%%%%%%%%%%%%%%%%%%%%%%%%%%%%%%%%%%%%%%%%%%%

\acknowledgments%%
We gratefully acknowledge support from NASA Astrophysics Theory Program under grant NNX17AK44G. Simulations in this paper made use of the {\tt REBOUND} code which can be downloaded freely at https://github.com/hannorein/rebound. 
This work utilized the RMACC Summit supercomputer, which is supported by the National Science Foundation (awards ACI-1532235 and ACI-1532236), the University of Colorado Boulder, and Colorado State University. The Summit supercomputer is a joint effort of the University of Colorado Boulder and Colorado State University.

%%%%%%%%%%%%%%%%%%%%%%%%%%%%%%%%%%%%%%%%%%%%%%%%%%%%%%%%%%%%%%%%%%%%%%%%%%%%%%%%

\bibliography{MasterRefs}

\begin{thebibliography}{}
\expandafter\ifx\csname natexlab\endcsname\relax\def\natexlab#1{#1}\fi

\bibitem[{{Alexander}(2017)}]{Ale17}
{Alexander}, T. 2017, \araa, 55, 17

\bibitem[{{Beloborodov} {et~al.}(1992){Beloborodov}, {Illarionov}, {Ivanov}, \&
  {Polnarev}}]{Bel92}
{Beloborodov}, A.~M., {Illarionov}, A.~F., {Ivanov}, P.~B., \& {Polnarev},
  A.~G. 1992, \mnras, 259, 209

\bibitem[{{Bender} {et~al.}(2005){Bender}, {Kormendy}, {Bower}, {Green},
  {Thomas}, {Danks}, {Gull}, {Hutchings}, {Joseph}, {Kaiser}, {Lauer},
  {Nelson}, {Richstone}, {Weistrop}, \& {Woodgate}}]{Bender2005}
{Bender}, R., {Kormendy}, J., {Bower}, G., {et~al.} 2005, \apj, 631, 280

\bibitem[{{Blaes} {et~al.}(2002){Blaes}, {Lee}, \& {Socrates}}]{Blaes2002}
{Blaes}, O., {Lee}, M.~H., \& {Socrates}, A. 2002, \apj, 578, 775

\bibitem[{{Bloom} {et~al.}(2011){Bloom}, {Giannios}, {Metzger}, {Cenko},
  {Perley}, {Butler}, {Tanvir}, {Levan}, {O'Brien}, {Strubbe}, {De Colle},
  {Ramirez-Ruiz}, {Lee}, {Nayakshin}, {Quataert}, {King}, {Cucchiara},
  {Guillochon}, {Bower}, {Fruchter}, {Morgan}, \& {van der Horst}}]{Bloom2011}
{Bloom}, J.~S., {Giannios}, D., {Metzger}, B.~D., {et~al.} 2011, Science, 333,
  203

\bibitem[{{Bogdanovi{\'c}} {et~al.}(2014){Bogdanovi{\'c}}, {Cheng}, \&
  {Amaro-Seoane}}]{Bog14}
{Bogdanovi{\'c}}, T., {Cheng}, R.~M., \& {Amaro-Seoane}, P. 2014, \apj, 788, 99

\bibitem[{{Bonnerot} \& {Rossi}(2018)}]{Bonnerot2018preprint}
{Bonnerot}, C., \& {Rossi}, E.~M. 2018, ArXiv e-prints, arXiv:1805.09329

\bibitem[{{Chen} {et~al.}(2011){Chen}, {Sesana}, {Madau}, \& {Liu}}]{Chen2011}
{Chen}, X., {Sesana}, A., {Madau}, P., \& {Liu}, F.~K. 2011, \apj, 729, 13

\bibitem[{{Cohn} \& {Kulsrud}(1978)}]{Coh78}
{Cohn}, H., \& {Kulsrud}, R.~M. 1978, \apj, 226, 1087

\bibitem[{{Couch} \& {Sharples}(1987)}]{Couch87}
{Couch}, W.~J., \& {Sharples}, R.~M. 1987, \mnras, 229, 423

\bibitem[{{Coughlin} {et~al.}(2017){Coughlin}, {Armitage}, {Nixon}, \&
  {Begelman}}]{Coughlin2017}
{Coughlin}, E.~R., {Armitage}, P.~J., {Nixon}, C., \& {Begelman}, M.~C. 2017,
  \mnras, 465, 3840

\bibitem[{{Dou} {et~al.}(2017){Dou}, {Wang}, {Yan}, {Jiang}, {Yang}, {Cutri},
  {Mainzer}, \& {Peng}}]{Dou2017}
{Dou}, L., {Wang}, T., {Yan}, L., {et~al.} 2017, \apjl, 841, L8

\bibitem[{{Einstein}(1916)}]{Ein16}
{Einstein}, A. 1916, Annalen der Physik, 354, 769

\bibitem[{{Ford} {et~al.}(2000){Ford}, {Kozinsky}, \& {Rasio}}]{Ford2000}
{Ford}, E.~B., {Kozinsky}, B., \& {Rasio}, F.~A. 2000, \apj, 535, 385

\bibitem[{{Frank} \& {Rees}(1976)}]{Fra76}
{Frank}, J., \& {Rees}, M.~J. 1976, \mnras, 176, 633

\bibitem[{{French} {et~al.}(2016){French}, {Arcavi}, \&
  {Zabludoff}}]{French2016}
{French}, K.~D., {Arcavi}, I., \& {Zabludoff}, A. 2016, \apjl, 818, L21

\bibitem[{{French} {et~al.}(2017){French}, {Arcavi}, \&
  {Zabludoff}}]{French2017}
---. 2017, \apj, 835, 176

\bibitem[{{Guillochon} \& {Ramirez-Ruiz}(2013)}]{Guillochon2013}
{Guillochon}, J., \& {Ramirez-Ruiz}, E. 2013, \apj, 767, 25

\bibitem[{{Hills}(1975)}]{Hil75b}
{Hills}, J.~G. 1975, \nat, 254, 295

\bibitem[{{Hopkins} \& {Quataert}(2010{\natexlab{a}})}]{Hop10a}
{Hopkins}, P.~F., \& {Quataert}, E. 2010{\natexlab{a}}, \mnras, 407, 1529

\bibitem[{{Hopkins} \& {Quataert}(2010{\natexlab{b}})}]{Hop10b}
---. 2010{\natexlab{b}}, \mnras, 405, L41

\bibitem[{{Ivanov} \& {Novikov}(2001)}]{Iva01}
{Ivanov}, P.~B., \& {Novikov}, I.~D. 2001, \apj, 549, 467

\bibitem[{{Kesden}(2012)}]{Kesden2012}
{Kesden}, M. 2012, \prd, 85, 024037

\bibitem[{{Lauer} {et~al.}(2005){Lauer}, {Faber}, {Gebhardt}, {Richstone},
  {Tremaine}, {Ajhar}, {Aller}, {Bender}, {Dressler}, {Filippenko}, {Green},
  {Grillmair}, {Ho}, {Kormendy}, {Magorrian}, {Pinkney}, \&
  {Siopis}}]{Lauer2005}
{Lauer}, T.~R., {Faber}, S.~M., {Gebhardt}, K., {et~al.} 2005, \aj, 129, 2138

\bibitem[{{Leloudas} {et~al.}(2016){Leloudas}, {Fraser}, {Stone}, {van Velzen},
  {Jonker}, {Arcavi}, {Fremling}, {Maund}, {Smartt}, {Kr{\`i}hler},
  {Miller-Jones}, {Vreeswijk}, {Gal-Yam}, {Mazzali}, {De Cia}, {Howell},
  {Inserra}, {Patat}, {de Ugarte Postigo}, {Yaron}, {Ashall}, {Bar},
  {Campbell}, {Chen}, {Childress}, {Elias-Rosa}, {Harmanen}, {Hosseinzadeh},
  {Johansson}, {Kangas}, {Kankare}, {Kim}, {Kuncarayakti}, {Lyman}, {Magee},
  {Maguire}, {Malesani}, {Mattila}, {McCully}, {Nicholl}, {Prentice},
  {Romero-Ca{\~n}izales}, {Schulze}, {Smith}, {Sollerman}, {Sullivan},
  {Tucker}, {Valenti}, {Wheeler}, \& {Young}}]{Leloudas2016}
{Leloudas}, G., {Fraser}, M., {Stone}, N.~C., {et~al.} 2016, Nature Astronomy,
  1, 0002

\bibitem[{{Lightman} \& {Shapiro}(1977)}]{Lig77}
{Lightman}, A.~P., \& {Shapiro}, S.~L. 1977, \apj, 211, 244

\bibitem[{{Lu} {et~al.}(2009){Lu}, {Ghez}, {Hornstein}, {Morris}, {Becklin}, \&
  {Matthews}}]{Lu2009}
{Lu}, J.~R., {Ghez}, A.~M., {Hornstein}, S.~D., {et~al.} 2009, \apj, 690, 1463

\bibitem[{{MacLeod} {et~al.}(2014){MacLeod}, {Goldstein}, {Ramirez-Ruiz},
  {Guillochon}, \& {Samsing}}]{Macleod2014}
{MacLeod}, M., {Goldstein}, J., {Ramirez-Ruiz}, E., {Guillochon}, J., \&
  {Samsing}, J. 2014, \apj, 794, 9

\bibitem[{{Madigan} {et~al.}(2018){Madigan}, {Halle}, {Moody}, {McCourt},
  {Nixon}, \& {Wernke}}]{Madigan2018}
{Madigan}, A.-M., {Halle}, A., {Moody}, M., {et~al.} 2018, \apj, 853, 141

\bibitem[{{Madigan} {et~al.}(2011){Madigan}, {Hopman}, \&
  {Levin}}]{Madigan2011}
{Madigan}, A.-M., {Hopman}, C., \& {Levin}, Y. 2011, \apj, 738, 99

\bibitem[{{Madigan} \& {McCourt}(2016)}]{Madigan2016}
{Madigan}, A.-M., \& {McCourt}, M. 2016, \mnras, 457, L89

\bibitem[{{Mainetti} {et~al.}(2017){Mainetti}, {Lupi}, {Campana}, {Colpi},
  {Coughlin}, {Guillochon}, \& {Ramirez-Ruiz}}]{Mai17}
{Mainetti}, D., {Lupi}, A., {Campana}, S., {et~al.} 2017, \aap, 600, A124

\bibitem[{{Maksym} {et~al.}(2013){Maksym}, {Ulmer}, {Eracleous}, {Guennou}, \&
  {Ho}}]{Maksym2013}
{Maksym}, W.~P., {Ulmer}, M.~P., {Eracleous}, M.~C., {Guennou}, L., \& {Ho},
  L.~C. 2013, \mnras, 435, 1904

\bibitem[{{Naoz} {et~al.}(2013){Naoz}, {Kocsis}, {Loeb}, \&
  {Yunes}}]{Naoz2013b}
{Naoz}, S., {Kocsis}, B., {Loeb}, A., \& {Yunes}, N. 2013, \apj, 773, 187

\bibitem[{{Poggianti}(2004)}]{Pog04}
{Poggianti}, B. 2004, in Baryons in Dark Matter Halos, ed. R.~{Dettmar},
  U.~{Klein}, \& P.~{Salucci}, 104

\bibitem[{{Press} \& {Teukolsky}(1977)}]{Pre77}
{Press}, W.~H., \& {Teukolsky}, S.~A. 1977, \apj, 213, 183

\bibitem[{{Rauch} \& {Ingalls}(1998)}]{Rau98}
{Rauch}, K.~P., \& {Ingalls}, B. 1998, \mnras, 299, 1231

\bibitem[{{Rauch} \& {Tremaine}(1996)}]{Rauch1996}
{Rauch}, K.~P., \& {Tremaine}, S. 1996, New Astronomy, 1, 149

\bibitem[{{Rees}(1988)}]{Ree88}
{Rees}, M.~J. 1988, \nat, 333, 523

\bibitem[{{Rein} \& {Liu}(2012)}]{Rein2012}
{Rein}, H., \& {Liu}, S.-F. 2012, \aap, 537, A128

\bibitem[{{Rein} \& {Spiegel}(2015)}]{Rein2015}
{Rein}, H., \& {Spiegel}, D.~S. 2015, \mnras, 446, 1424

\bibitem[{{Saxton} {et~al.}(2016){Saxton}, {Younsi}, \& {Wu}}]{Sax16}
{Saxton}, C.~J., {Younsi}, Z., \& {Wu}, K. 2016, \mnras, 461, 4295

\bibitem[{{Shapiro} \& {Marchant}(1978)}]{Sha78}
{Shapiro}, S.~L., \& {Marchant}, A.~B. 1978, \apj, 225, 603

\bibitem[{{Sil'chenko} {et~al.}(1998){Sil'chenko}, {Burenkov}, \&
  {Vlasyuk}}]{Sil98}
{Sil'chenko}, O.~K., {Burenkov}, A.~N., \& {Vlasyuk}, V.~V. 1998, \aap, 337,
  349

\bibitem[{{Stone} {et~al.}(2017){Stone}, {Generozov}, {Vasiliev}, \&
  {Metzger}}]{Stone2017}
{Stone}, N.~C., {Generozov}, A., {Vasiliev}, E., \& {Metzger}, B.~D. 2017,
  ArXiv e-prints, arXiv:1709.00423

\bibitem[{Stone \& Metzger(2016)}]{Stone2016a}
Stone, N.~C., \& Metzger, B.~D. 2016, 455, 859

\bibitem[{{Stone} \& {van Velzen}(2016)}]{Stone2016b}
{Stone}, N.~C., \& {van Velzen}, S. 2016, \apjl, 825, L14

\bibitem[{{Tadhunter} {et~al.}(2017){Tadhunter}, {Spence}, {Rose}, {Mullaney},
  \& {Crowther}}]{Tadhunter2017}
{Tadhunter}, C., {Spence}, R., {Rose}, M., {Mullaney}, J., \& {Crowther}, P.
  2017, Nature Astronomy, 1, 0061

\bibitem[{{Tremaine}(1995)}]{Tremaine1995}
{Tremaine}, S. 1995, \aj, 110, 628

\bibitem[{{van Velzen}(2018)}]{vanVelzen2018}
{van Velzen}, S. 2018, \apj, 852, 72

\bibitem[{{van Velzen} {et~al.}(2011){van Velzen}, {Farrar}, {Gezari},
  {Morrell}, {Zaritsky}, {{\"O}stman}, {Smith}, {Gelfand}, \&
  {Drake}}]{vanVelzen2011}
{van Velzen}, S., {Farrar}, G.~R., {Gezari}, S., {et~al.} 2011, \apj, 741, 73

\bibitem[{{van Velzen} {et~al.}(2016){van Velzen}, {Anderson}, {Stone},
  {Fraser}, {Wevers}, {Metzger}, {Jonker}, {van der Horst}, {Staley}, {Mendez},
  {Miller-Jones}, {Hodgkin}, {Campbell}, \& {Fender}}]{vanVelzen2016}
{van Velzen}, S., {Anderson}, G.~E., {Stone}, N.~C., {et~al.} 2016, Science,
  351, 62

\bibitem[{{Wang} \& {Merritt}(2004)}]{Wan04}
{Wang}, J., \& {Merritt}, D. 2004, \apj, 600, 149

\bibitem[{{Zauderer} {et~al.}(2011){Zauderer}, {Berger}, {Soderberg}, {Loeb},
  {Narayan}, {Frail}, {Petitpas}, {Brunthaler}, {Chornock}, {Carpenter},
  {Pooley}, {Mooley}, {Kulkarni}, {Margutti}, {Fox}, {Nakar}, {Patel},
  {Volgenau}, {Culverhouse}, {Bietenholz}, {Rupen}, {Max-Moerbeck}, {Readhead},
  {Richards}, {Shepherd}, {Storm}, \& {Hull}}]{Zauderer2011}
{Zauderer}, B.~A., {Berger}, E., {Soderberg}, A.~M., {et~al.} 2011, \nat, 476,
  425

\end{thebibliography}

\end{document}